\documentclass[twocolumn,epjc3,final]{svjour3}
\smartqed  
\RequirePackage{graphicx}
\RequirePackage{mathptmx}      
\RequirePackage{latexsym,amssymb}   
\RequirePackage[numbers,sort&compress]{natbib}
\RequirePackage[colorlinks,citecolor=blue,urlcolor=blue,linkcolor=blue]{hyperref} 
\journalname{Eur. Phys. J. Plus}

\def\prd{Physical Review D }
\def\aap{Astron. Astroph.}

\def\apjl{Astroph. J. }
\def\mnras{Mon. Not. R. Astron. Soc.}
\def\nar{New Astron. Rev.}
\def\nat{Nature}
\def\jcap{JCAP}

\begin{document}

\title{Testing the Weak Equivalence Principle and Lorentz Invariance with Multiwavelength Polarization Observations of GRB Optical Afterglows}



\author{Jun-Jie Wei \thanksref{addr1,addr2,e1}
	\and
	Xue-Feng Wu\thanksref{addr1,addr2,e2}
}
\thankstext{e1}{e-mail: jjwei@pmo.ac.cn}
\thankstext{e2}{e-mail: xfwu@pmo.ac.cn}
\institute{Purple Mountain Observatory, Chinese Academy of Sciences, Nanjing 210023, China \label{addr1}
\and School of Astronomy and Space Sciences, University of Science and Technology of China, Hefei 230026, China \label{addr2} }

\date{Received: date / Accepted: date}

\maketitle

\begin{abstract}
Violations of both the weak equivalence principle (WEP) and Lorentz invariance can produce vacuum birefringence,
which leads to an energy-dependent rotation of the polarization vector of linearly polarized emission from a
given astrophysical source. However, the search for the birefringent effect has been hindered by our ignorance
concerning the intrinsic polarization angle in different energy bands. Considering the contributions to the
observed linear polarization angle from both the intrinsic polarization angle and the rotation angles induced
by violations of the WEP and Lorentz invariance, and assuming the intrinsic polarization angle is an unknown
constant, we simultaneously obtain robust bounds on possible deviations from the WEP and Lorentz invariance,
by directly fitting the multiwavelength polarimetric data of the optical afterglows of gamma-ray burst (GRB)
020813 and GRB 021004. Here we show that at the $3\sigma$ confidence level, the difference of the parameterized
post-Newtonian parameter $\gamma$ values characterizing the departure from the WEP is constrained to be
$\Delta\gamma=\left(-4.5^{+10.0}_{-16.0}\right)\times10^{-24}$ and the birefringent parameter $\eta$ quantifying
the broken degree of Lorentz invariance is limited to be $\eta=\left(6.5^{+15.0}_{-14.0}\right)\times10^{-7}$.
These are the first simultaneous verifications of the WEP and Lorentz invariance in the photon sector.
More stringent limits can be expected as the analysis presented here is applied to future multiwavelength
polarization observations in the prompt gamma-ray emission of GRBs.
\end{abstract}

\section{Introduction}\label{sec:intro}

The weak equivalence principle (WEP) is a fundamental postulate of general relativity as well as of
many other metric theories of gravity. One statement of the WEP is that the trajectory of any freely
falling, uncharged test body does not depend on its internal structure and composition \cite{2006LRR.....9....3W,2014LRR....17....4W}.
It implies that different species of messenger particles (e.g., photons, neutrinos, or
gravitational waves), or the same species of particles but with different internal structures (e.g,
energies or polarization states), if radiated simultaneously from the same astrophysical source and
passing through the same gravitational field, should arrive at our Earth at the same time.
The WEP test can therefore be performed by comparing the arrival-time differences between correlated
particles from the same astrophysical source (e.g., \cite{1988PhRvL..60..176K,1988PhRvL..60..173L,2015ApJ...810..121G,2015PhRvL.115z1101W,
2016ApJ...818L...2W,2016JCAP...08..031W,2017JCAP...11..035W,2019JHEAp..22....1W,
2016PhLB..756..265K,2016ApJ...827...75L,2016ApJ...821L...2N,2016MNRAS.460.2282S,
2016ApJ...820L..31T,2016PhRvL.116o1101W,2016PhRvD..94b4061W,2017PhRvD..95j3004W,
2016PhRvD..94j1501Y,2017ApJ...848L..13A,2017PhLB..770....8L,2017ApJ...837..134Z,
2018EPJC...78...86D,2018ApJ...861...66L,2018PhRvD..97h3013S,2018ApJ...860..173Y,
2019EPJC...79..185B,2019PhRvD.100j3002L,2019ApJ...882L..13X}).
Additionally, if the WEP is invalid then arrival times of photons with right- and left-handed circular
polarizations should differ slightly, leading to a frequency-dependent rotation of the polarization
plane of a linearly polarized light. Thus, polarimetric observations of astrophysical sources can
also be used to test the WEP \cite{2017MNRAS.469L..36Y,2019PhRvD..99j3012W,2020MNRAS.493.1782Y}. Currently, the best
upper limit on a deviation from the WEP has been obtained from the gamma-ray polarization measurement
of gamma-ray burst (GRB) 061122 \cite{2019PhRvD..99j3012W}. The WEP passes this extraordinarily
stringent test with an accuracy of $\mathcal{O}(10^{-33})$.

Lorentz invariance is a foundational symmetry of Einstein's theory of relativity. However, many
quantum gravity theories seeking to unify quantum mechanics and general relativity predict that
Lorentz invariance may be broken at the Planck energy scale $E_{\rm Pl}\simeq1.22\times10^{19}$ GeV
\cite{1989PhRvD..39..683K,1991NuPhB.359..545K,1995PhRvD..51.3923K,1998Natur.393..763A,
2005LRR.....8....5M,2005hep.ph....6054B,2013LRR....16....5A,2014RPPh...77f2901T}.
As a consequence of Lorentz invariance violation (LIV), the polarization vector of linearly polarized
photons would make an energy-dependent rotation, also known as vacuum birefringence.
Lorentz invariance can therefore be tested with astrophysical polarization measurements
(e.g.,
\cite{1999PhRvD..59l4021G,
2001PhRvD..64h3007G,
2001PhRvL..87y1304K,
2006PhRvL..97n0401K,
2007PhRvL..99a1601K,
2008ApJ...689L...1K,
2013PhRvL.110t1601K,
2003Natur.426Q.139M,
2004PhRvL..93b1101J,
2007MNRAS.376.1857F,
2009JCAP...08..021G,
2011PhRvD..83l1301L,
2011APh....35...95S,
2012PhRvL.109x1104T,
2013MNRAS.431.3550G,
2014MNRAS.444.2776G,
2016MNRAS.463..375L,
2017PhRvD..95h3013K,
2019PhRvD..99c5045F,
2019MNRAS.485.2401W}).
The presence of linear polarization in the prompt gamma-ray emission of GRBs sets the strictest upper
limit to date on the birefringent parameter, namely $\eta<\mathcal{O}(10^{-16})$ \cite{2012PhRvL.109x1104T,2013MNRAS.431.3550G,2014MNRAS.444.2776G,2016MNRAS.463..375L,2019MNRAS.485.2401W}
(see also \cite{2011RvMP...83...11K,2013CQGra..30m3001L} and summary constraints for LIV therein).

In general, it is hard to know the intrinsic polarization angles for photons with different energies
from a given source. If one possesses this information, a rotation angle of the polarization plane,
which induced by astrophysical effects (e.g., the WEP violation or LIV), could be
directly extracted by measuring the difference between the known intrinsic polarization angle and
the observed polarization angle for photons at a certain energy. Even in the absence of such knowledge,
however, birefringent effects can still be constrained for sources at arbitrary redshifts. The reason
is as follows \cite{2012PhRvL.109x1104T}. It is believed that if the rotation angle (denoted by
$\Delta\phi$) differs by more than $\pi/2$ over an energy range $[E_{1},\;E_{2}]$, then the net
polarization of the signal would be substantially depleted and could not be as high as the observed level.
That is, the detection of high polarization means that the relative rotation angle
$|\Delta\phi(E_{2})-\Delta\phi(E_{1})|$ should not be too large. Therefore, some upper limits on
violations of the WEP and Lorentz invariance can be obtained under the assumption that
$|\Delta\phi(E_{2})-\Delta\phi(E_{1})|$ is smaller than $\pi/2$. However, through the detailed analyses
for the evolution of GRB polarization arising from violations of the WEP and Lorentz invariance,
Lin et al. \cite{2016MNRAS.463..375L} and Wei \& Wu \cite{2019PhRvD..99j3012W} proved that more than 60\% of the initial
polarization can be conserved even if $|\Delta\phi(E_{2})-\Delta\phi(E_{1})|$ is as large as $\pi/2$.
This is conflict with the intuition that $|\Delta\phi(E_{2})-\Delta\phi(E_{1})|$ could not be larger
than $\pi/2$ when high polarization is detected. Hence, it is inappropriate to simply use $\pi/2$ as
the upper limit of $|\Delta\phi(E_{2})-\Delta\phi(E_{1})|$ to constrain deviations from the WEP and
Lorentz invariance. Furthermore, even though some upper limits of the violations were found to be
extremely small \cite{2012PhRvL.109x1104T,2013MNRAS.431.3550G,2014MNRAS.444.2776G,2016MNRAS.463..375L,
2017MNRAS.469L..36Y,2019MNRAS.485.2401W,2019PhRvD..99j3012W}, the outcomes of these limits are lack of
significantly statistical robustness.

In this work, we propose that an intrinsic polarization angle can be extracted and a more robust
bound on a deviation from the WEP or from Lorentz invariance can be derived as well, by directly fitting
the multiwavelength polarization observations of astrophysical sources. More importantly,
the analysis of the multiwavelength polarimetric data also allows us to simultaneously test
the WEP and Lorentz invariance, when we consider that the rotation angle is caused by violations of
both the WEP and Lorentz invariance.

\section{Tests of the Weak Equivalence Principle}
\label{sec:WEP}

Adopting the parameterized post-Newtonian (PPN) formalism, the time interval required for test particles to
travel across a given distance would be longer in the presence of a gravitational potential $U(r)$ by
\begin{equation}
t_{\rm gra}=-\frac{1+\gamma}{c^3}\int_{r_e}^{r_o}~U(r)dr\;,
\end{equation}
where $\gamma$ is one of the PPN parameters ($\gamma$ reflects the level of space curved by unit rest
mass) and the integration is along the propagation path from the emitting source $r_e$ to the observer $r_o$.
This effect is known as Shapiro time delay \cite{1964PhRvL..13..789S}. It is important to note that all metric theories of
gravity satisfying the WEP predict that any two test particles traveling in the same gravitational
field must follow the same trajectory and undergo the identical Shapiro delay. In other words, as long
as the WEP is valid, all metric theories predict that the measured value of $\gamma$ should be the
same for all test particles \cite{2006LRR.....9....3W,2014LRR....17....4W}. The accuracy of the WEP
can therefore be characterized by placing constraints on the differences of the $\gamma$ values for
different particles.

Linearly polarized light is a superposition of two monochromatic waves with opposite circular polarizations
(labeled with $r$ and $l$). If the WEP is broken, different $\gamma$ values might be measured with right-
and left-handed circularly polarized photons, leading to the slight arrival-time difference of these two
circular components. The arrival-time lag is then given by
\begin{equation}
\Delta t_{\rm gra}=\left|\frac{\Delta\gamma}{c^3}\int_{r_e}^{r_o}~U(r)dr\right|\;,
\label{eq:delta-tgra}
\end{equation}
where $\Delta\gamma=\gamma_{r}-\gamma_{l}$ corresponds to the difference of the $\gamma$ values for different circular polarization states.
To compute $\Delta t_{\rm gra}$ with Equation~(\ref{eq:delta-tgra}), we have to figure out the gravitational
potential $U(r)$. For a cosmic source, $U(r)$ should have contributions from the gravitational potentials of
the Milky Way $U_{\rm MW}(r)$, the intergalactic space $U_{\rm IG}(r)$, and the source host galaxy $U_{\rm host}(r)$.
Since the potential models of $U_{\rm IG}(r)$ and $U_{\rm host}(r)$ are poorly understood, for the purposes
of obtaining conservative limits, we here just consider the Milky Way gravitational potential. Adopting a
Keplerian potential\footnote{Although the potential model of the Milky Way is still not well known,
ref.~\cite{1988PhRvL..60..176K} examined two popular potential models (i.e., the Keplerian potential and the isothermal
potential) and suggested that the adoption of a different model for $U_{\rm MW}(r)$ has only a minimal influence
on the WEP tests.} $U(r)=-GM/r$ for the Milky Way, we thus have \cite{1988PhRvL..60..173L,2016PhRvD..94b4061W}
\begin{eqnarray}
\Delta t_{\rm gra}= \Delta\gamma \frac{GM_{G}}{c^{3}} \times \qquad\qquad\qquad\qquad\qquad\qquad\qquad\\ \nonumber
\ln \left\{ \frac{ \left[d+\left(d^{2}-b^{2}\right)^{1/2}\right] \left[r_{G}+s_{\rm n}\left(r_{G}^{2}-b^{2}\right)^{1/2}\right] }{b^{2}} \right\}\;,
\label{eq:gammadiff}
\end{eqnarray}
where $M_{G}\simeq6\times10^{11}M_{\odot}$ is the mass of the Milky Way \cite{2012ApJ...761...98K}, $d$ is
approximated as the distance from the source to our Earth, $b$ represents the impact parameter of the light paths
relative to our Galactic center, and $r_{G}=8.3$ kpc denotes the distance of our Galactic center. Here we use
$s_{\rm n}=+1$ or $s_{\rm n}=-1$ to correspond to the cases where the source is located along the Galactic center
or anti-Galactic center. The impact parameter $b$ can be estimated as
\begin{equation}
b=r_{G}\sqrt{1-(\sin \delta_{s} \sin \delta_{G}+\cos \delta_{s} \cos \delta_{G} \cos(\beta_{s}-\beta_{G}))^{2}}\;,
\label{eq:b}
\end{equation}
where $\beta_{s}$ and $\delta_{s}$ are the right ascension and declination of the source in the equatorial coordinate
system and ($\beta_{G}=17^{\rm h}45^{\rm m}40.04^{\rm s}$, $\delta_{G}=-29^{\circ}00^{'}28.1^{''}$) represent
the coordinates of the Galactic center \cite{2009ApJ...692.1075G}.

As mentioned above, a possible violation of the WEP can lead to the slight arrival-time difference of photons with
right- and left-handed circular polarizations. In this case, the polarization vector of a linearly polarized light
will rotate during the propagation. The rotation angle induced by the WEP violation is expressed as \cite{2017MNRAS.469L..36Y,2019PhRvD..99j3012W}
\begin{equation}
\Delta\phi_{\rm WEP}\left(E\right)=\Delta t_{\rm gra}\frac{2\pi c}{\lambda}=\Delta t_{\rm gra}\frac{E}{\hbar}\;,
\label{eq:theta-WEP}
\end{equation}
where $E$ is the observed photon energy.

If the birefringent effect arising from the WEP violation is considered here,
the observed linear polarization angle ($\phi_{\rm obs}$) for photons emitted with energy $E$ from an astrophysical
source should consist of two terms
\begin{equation}
\phi_{\rm obs}=\phi_{0}+\Delta\phi_{\rm WEP}\left(E\right)\;,
\end{equation}
where $\phi_{0}$ represents the intrinsic polarization angle. As $\phi_{0}$ is unknown, the exact value of
$\Delta\phi_{\rm WEP}$ is not available. Yet, an upper limit on the $\gamma$ discrepancy ($\Delta\gamma$)
can be obtained by setting the upper limit of the relative rotation angle $|\Delta\phi_{\rm WEP}(E_{2})-\Delta\phi_{\rm WEP}(E_{1})|$
to be $\pi/2$, which is based on the argument that the observed polarization degree will be significantly
suppressed if $|\Delta\phi_{\rm WEP}(E_{2})-\Delta\phi_{\rm WEP}(E_{1})|>\pi/2$ over an observed energy range $[E_{1},\;E_{2}]$,
regardless of the intrinsic polarization fraction at the corresponding rest-frame energy range \cite{2017MNRAS.469L..36Y}.
Instead of requiring the more complicated and indirect argument, here we simply assume that all photons
in the observed bandpass are emitted with the same (unknown) intrinsic polarization angle. In this case,
we expect to observe the birefringent effect induced by the WEP violation as an energy-dependent linear
polarization vector. Such an observation could confirm the existence of a birefringent effect and give
a robust limit on the WEP violation. We look for a similar energy-dependent trend in multiwavelength
polarization observations of present astrophysical sources. One can see from Equation~(\ref{eq:theta-WEP})
that much more stringent constraints on $\Delta\gamma$ can be obtained from the higher energy band of
polarization observations. Unfortunately, there are not existing multiwavelength polarization observations
(i.e., more than three wavelength bins) in the gamma-ray or X-ray energy band. We explore here the implications
and limits that can be set by multiwavelength linear polarization observations from the optical afterglows
of GRB 020813 \cite{2003ApJ...584L..47B} and GRB 021004 \cite{2003A&A...410..823L}.

\begin{figure}
\vskip-0.2in
\centerline{\includegraphics[angle=0,width=1.1\hsize]{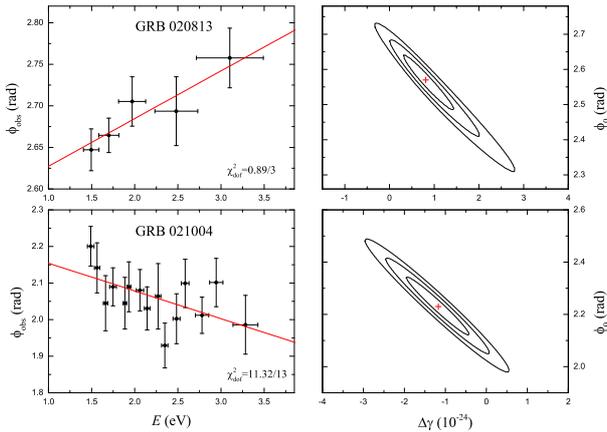}}
\vskip-0.3in
\caption{Fit to multiwavelength polarimetric observations of the optical afterglows of GRB 020813
(upper panels; the polarization angles are time-averaged) and GRB 021004 (lower panels).
Left panels: observed polarization angle $\phi_{\rm obs}$ as a function of the energy $E$,
and the best-fit theoretical curves for the case of the WEP violation.
Right panels: 1--3$\sigma$ confidence levels in the $\Delta\gamma$--$\phi_{0}$ plane.
The plus symbol represents the best fit, corresponding to the reduced chi-square value $\chi^{2}_{\rm dof}$.}
\label{fig:f1}
\end{figure}

GRB 020813 was detected by the \emph{High Energy Transient Explore 2} (\emph{HETE2}) on 13 August 2002, with
coordinates R.A.=$19^{\rm h}46^{\rm m}38^{\rm s}$ and Dec.=$-19^{\circ}35^{'}16^{''}$ \cite{2002GCN..1471....1V}.
Its redshift has been measured to be $z=1.255$ \cite{2003ApJ...584L..47B}. The multiwavelength
[(4000, 5000, 6300, 7300, 8300)$\pm$500 {\AA}] polarization measurements of the optical afterglow of GRB 020813
were carried out during 4.7--7.9 hr after the burst. The observed linear polarization is in the range of
1.8\%--2.4\% (see Table 2 of ref.~\cite{2003ApJ...584L..47B}). At $t\sim7.36$ hr after the burst, the observed
polarization angles in five wavelength bins are $153^{\circ}\pm1^{\circ}$, $149^{\circ}\pm1^{\circ}$,
$156^{\circ}\pm1^{\circ}$, $153^{\circ}\pm1^{\circ}$, and $149^{\circ}\pm1^{\circ}$, respectively.
At $t\sim6.27$ hr, the corresponding polarization angles are $160^{\circ}\pm1^{\circ}$, $155^{\circ}\pm1^{\circ}$,
$151^{\circ}\pm1^{\circ}$, $150^{\circ}\pm1^{\circ}$, and $151^{\circ}\pm2^{\circ}$, respectively.
At $t\sim5.16$ hr, the corresponding polarization angles are $161^{\circ}\pm1^{\circ}$, $159^{\circ}\pm1^{\circ}$,
$158^{\circ}\pm1^{\circ}$, $155^{\circ}\pm1^{\circ}$, and $155^{\circ}\pm1^{\circ}$, respectively.
We allow a temporal variation of the polarization and consider only the relative polarization angle.
In each wavelength bin, the time-averaged polarization angle during three observational periods
is calculated through $\overline{\phi}_{\rm obs}=\sum_{i}\phi_{{\rm obs},i}/3$.
The scatters in the shift between $\phi_{{\rm obs},i}$ and $\overline{\phi}_{\rm obs}$, i.e.,
$\sigma_{\phi_{i}}=|\phi_{{\rm obs},i}-\overline{\phi}_{\rm obs}|$, provide an estimate of the error in $\overline{\phi}_{\rm obs}$,
i.e., $\sigma_{\overline{\phi}}=(\sum_{i}\sigma^{2}_{\phi_{i}})^{1/2}/3$. The time-averaged polarization angle
as a function of energy is displayed in the upper-left panel of Figure~\ref{fig:f1}. We fit the $\gamma$ discrepancy
($\Delta\gamma$), along with the intrinsic polarization angle $\phi_{0}$, by maximizing the likelihood
function:
\begin{equation}
 \mathcal{L} = \prod_{i}\frac{1}{\sqrt{2\pi}\,\sigma_{{\rm tot}, i}}\times
             \exp\left[-\frac{\left(\phi_{{\rm obs},i}-\phi_{\rm th}\left(E_{i}\right)\right)^{2}}{2\sigma^{2}_{{\rm tot}, i}}\right]\;,
\label{eq:likelihood}
\end{equation}
where $\phi_{\rm th}\left(E_{i}\right)=\phi_{0}+\Delta\phi_{\rm WEP}\left(\Delta\gamma,\;E_{i}\right)$ and
the variance
\begin{equation}
\sigma_{{\rm tot},i}^{2}=\sigma_{\phi_{{\rm obs},i}}^{2}+\left(\frac{\Delta\phi_{\rm WEP}}{E_{i}}\sigma_{E_{i}}\right)^{2}
\end{equation}
is given in terms of the measurement error $\sigma_{\phi_{{\rm obs},i}}$ in $\phi_{{\rm obs},i}$ and the
propagated error of $\sigma_{E_{i}}$. The 1--3$\sigma$ confidence levels in the $\Delta\gamma$--$\phi_{0}$ plane
are presented in the upper-right panel of Figure~\ref{fig:f1}. The best-fitting values are $\Delta\gamma=0.81\times10^{-24}$
and $\phi_{0}=2.57$ rad. Our constraints show that $\Delta\gamma$ is consistent with $0$ at the $2.5\sigma$ confidence level,
implying that there is no convincing evidence for the violation of the WEP. At the $3\sigma$ confidence level, the limits on
$\Delta\gamma$ are $-1.6\times10^{-25}<\Delta\gamma<2.7\times10^{-24}$.

\begin{figure}
\vskip-0.2in
\centerline{\includegraphics[angle=0,width=1.0\hsize]{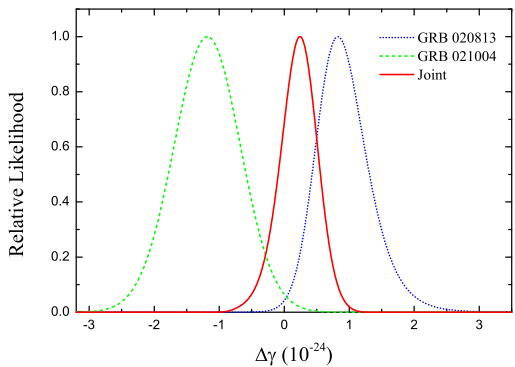}}
\vskip-0.1in
\caption{Individual and joint constraints on the difference of the $\gamma$ values
from the optical polarimetry of GRB 020813 and GRB 021004.}
\label{fig:f2}
\end{figure}

GRB 021004 was detected by $HETE2$ on 4 October 2002, with coordinates R.A.=$00^{\rm h}26^{\rm m}57^{\rm s}$
and Dec.=$+18^{\circ}55^{'}44^{''}$ \cite{2002GCN..1565....1S}. Its redshift is $z=2.328$ \cite{2002GCN..1618....1M}.
Here we directly take the reduced spectropolarimetric observations for the optical counterpart to GRB 021004
that presented in ref.~\cite{2003A&A...410..823L}. As illustrated in the lower-left panel of Figure~\ref{fig:f1},
the observed polarization angle has a negative dependence on the energy, rather than a positive dependence.
The parameter constraints are shown in the lower-right panel of Figure~\ref{fig:f1}. We see here that
the best-fit corresponds to $\Delta\gamma=-1.17\times10^{-24}$ and $\phi_{0}=2.23$ rad. The data set is consistent
with the possibility of no WEP violation at all (i.e., $\Delta\gamma=0$) at the $2.4\sigma$ confidence level.
At the $3\sigma$ confidence level, we have $-2.7\times10^{-24}<\Delta\gamma<3.1\times10^{-25}$.

Now we combine the polarimetric data of GRB 020813 and GRB 021004 to further investigate the possible birefringent
effect that arises from the WEP violation. The calculation procedure is as follows. For each data set,
we first derive the 1-D probability distribution of $\Delta\gamma$ by marginalizing over the intrinsic polarization
angle $\phi_{0}$. The joint probability of each $\Delta\gamma$ for two bursts is then calculated with the total
likelihood function
$\mathcal{L}_{\rm tot}(\Delta\gamma_{i})\propto\mathcal{L}_{\rm 020813}(\Delta\gamma_{i})\cdot \mathcal{L}_{\rm 021004}(\Delta\gamma_{i})$,
where $i$ indicates the $i$th $\Delta\gamma$. In Figure~\ref{fig:f2}, we present the marginalized likelihood
distributions of $\Delta\gamma$ derived from the optical polarimetry of GRB 020813 (dotted curve), GRB 021004
(dashed curve), and their combination (solid curve), respectively. At the $3\sigma$ confidence level,
the combined constraint on $\Delta\gamma$ is $\left(0.2^{+0.8}_{-1.0}\right)\times10^{-24}$.

Under the same assumption that the Shapiro delay is attributed to the Milky Way's gravity,
ref. \cite{2017MNRAS.469L..36Y} obtained the current best limit of $\Delta\gamma<1.6\times10^{-27}$
from the gamma-ray polarimetric data of GRB 110721A. While our combined limit is three orders of magnitude
less precise, our analysis, which may be unique in the literature, does constrain $\Delta\gamma$ by directly
fitting the multiwavelength polarimetric observations in the optical band. As such, our analysis provides
an independent test of the WEP and has the promise to compliment existing WEP tests.

\section{Constraints on the Violation of Lorentz Invariance}
\label{sec:LIV}

In the photon sector, the Lorentz-violating dispersion relation can be expressed as \cite{2003PhRvL..90u1601M}
\begin{equation}\label{eq:dispersion}
  E_{\pm}^2=p^2c^2\pm \frac{2\eta}{E_{\rm pl}} p^3c^3\;,
\end{equation}
where $E_{\rm pl}\approx 1.22\times 10^{19}$ GeV is the Planck energy, $\pm$ denotes the left- or right-handed
circular polarization states, and $\eta$ is a dimensionless parameter characterizing the broken degree of Lorentz invariance.
If $\eta\neq0$, then group velocities for different circular polarization states should differ slightly, leading
to vacuum birefringence and a phase rotation of linear polarization \cite{1999PhRvD..59l4021G,2001PhRvD..64h3007G,
2003Natur.426Q.139M,2004PhRvL..93b1101J}. The rotation angle of the polarization vector for linearly polarized photons
propagating from the source at redshift $z$ to the observer is given by \cite{2011PhRvD..83l1301L,2012PhRvL.109x1104T}
\begin{equation}\label{eq:theta-LIV}
  \Delta\phi_{\rm LIV}(E)=\eta\frac{E^2}{\hbar E_{\rm pl}}\int_0^z\frac{1+z'}{H(z')}dz'\;,
\end{equation}
where $E$ is the energy of the observed light. Also, $H(z)=H_0\left[\Omega_{\rm m}(1+z)^3+\Omega_{\Lambda}\right]^{1/2}$
is the Hubble parameter at $z$, where the standard flat $\Lambda$CDM model with parameters
$H_{0}=67.36$ km $\rm s^{-1}$ $\rm Mpc^{-1}$, $\Omega_{\rm m}=0.315$, and $\Omega_{\Lambda}=1-\Omega_{\rm m}$ is adopted
\cite{2018arXiv180706209P}.

\begin{figure}
\vskip-0.2in
\centerline{\includegraphics[angle=0,width=1.1\hsize]{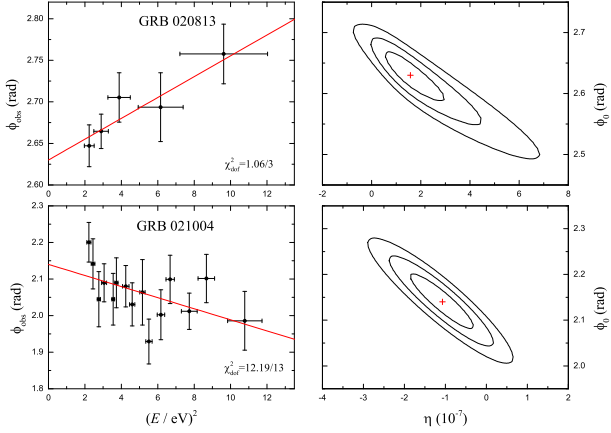}}
\vskip-0.3in
\caption{Similar to Figure~\ref{fig:f1}, but for the case of LIV.}
\label{fig:f3}
\end{figure}

\begin{figure}
\vskip-0.2in
\centerline{\includegraphics[angle=0,width=1.0\hsize]{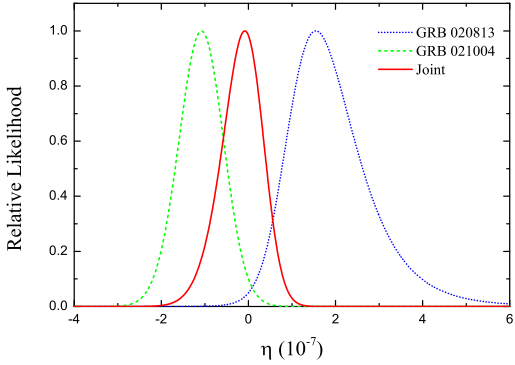}}
\vskip-0.1in
\caption{Individual and joint constraints on the birefringent parameter $\eta$
from the optical polarimetry of GRB 020813 and GRB 021004.}
\label{fig:f4}
\end{figure}

If the birefringent effect arising from LIV is considered here, the observed polarization angle for photons
at a certain energy $E$ with an intrinsic polarization angle $\phi_{0}$ should be
\begin{equation}
\phi_{\rm obs}=\phi_{0}+\Delta\phi_{\rm LIV}\left(E\right)\;.
\end{equation}
The observed polarization angles of GRB 020813 and GRB 021004 as a function of $E^{2}$ are shown in the left panels
of Figure~\ref{fig:f3}. To find the best-fitting birefringent parameter $\eta$ and the intrinsic polarization angle
$\phi_{0}$, we also adopt the method of maximum likelihood estimation. The adopted likelihood function is the same as
Equation~(\ref{eq:likelihood}), except now $\phi_{\rm th}\left(E_{i}\right)=\phi_{0}+\Delta\phi_{\rm LIV}\left(\eta,\;E_{i}\right)$ and
$\sigma_{{\rm tot},i}^{2}=\sigma_{\phi_{{\rm obs},i}}^{2}+\left(2\frac{\Delta\phi_{\rm LIV}}{E_{i}}\sigma_{E_{i}}\right)^{2}$.
The resulting constraints on $\eta$ and $\phi_{0}$ are displayed in the right panels of Figure~\ref{fig:f3}.
For GRB 020813, the best-fitting parameters are $\eta=1.58\times10^{-7}$ and $\phi_{0}=2.63$ rad, and $\eta$
is consistent with 0 (i.e., no evidence of LIV) at the $2.5\sigma$ confidence level.
For GRB 021004, the best-fit corresponds to $\eta=-1.07\times10^{-7}$ and $\phi_{0}=2.14$ rad, and the data set
is consistent with the possibility of no LIV at all (i.e., $\eta=0$) at the $2.2\sigma$ confidence level.
The 1-D marginalized likelihood distributions of $\eta$ derived from the optical polarimetry of GRB 020813 (dotted curve),
GRB 021004 (dashed curve), and their combination (solid curve) are plotted in Figure~\ref{fig:f4}.
We find that the $3\sigma$ level joint-constraint is $\eta=\left(-0.1^{+1.2}_{-1.7}\right)\times10^{-7}$,
which is in good agreement with the result of ref.~\cite{2007MNRAS.376.1857F}.

Using the detections of prompt emission polarization in GRBs, ref.~\cite{2019MNRAS.485.2401W}
set the hitherto most stringent constraint on the birefringent parameter, i.e., $\eta<\mathcal{O}(10^{-16})$
(see also \cite{2013MNRAS.431.3550G,2014MNRAS.444.2776G,2016MNRAS.463..375L}).
While our optical polarization constraint is not competitive with observations of gamma-ray polarization,
there is merit to the result. We use the spectropolarimetric data in order to directly
measure the energy-dependent change of the polarization angle. This is an improvement over many previous analyses,
which made use of the argument that birefringence would significantly reduce polarization over a broad bandwidth
to obtain limits on the birefringent parameter $\eta$.

\section{Limits on violations of both the WEP and Lorentz Invariance}
\label{sec:both}
The Einstein equivalence principle entails three assumptions:
the universality of free fall (WEP), local Lorentz invariance, and
local position invariance of non-gravitational experiments
\cite{2006LRR.....9....3W,2014LRR....17....4W}. Indeed, it is well
known that particles endowed with modified dispersion relation like
Equation~(\ref{eq:dispersion}) are not only violating Lorentz invariance
but also the WEP as in general they will not follow the geodesic of any
metric. And actually to describe such modified dispersion relations,
one has to assume the presence of an extra field beyond the metric
such as an aether vector field, see e.g. Einstein-{\AE}ther gravity theory
\cite{2001PhRvD..64b4028J,2004PhRvD..70b4003J,2004PhRvD..69f4005E,2006PhRvD..73f4015F}.

Since both the WEP violation and the LIV effect can lead to an energy-dependent rotation of the linear polarization
angle, the observed polarization angle
\begin{equation}
\phi_{\rm obs}=\phi_{0}+\Delta\phi_{\rm WEP}\left(E\right)+\Delta\phi_{\rm LIV}\left(E\right)
\end{equation}
in principle should have contributions from the intrinsic polarization angle and the rotation angles induced by
violations of the WEP and Lorentz invariance, respectively. If we suppose that the energy-dependent rotation angle
is attributed to these two causes, the difference of the $\gamma$ values and the birefringent parameter $\eta$
can be simultaneously constrained by fitting the multiwavelength polarimetric observations of the optical afterglows
of GRB 020813 and GRB 021004. Similarly, we maximize the likelihood function (Equation~(\ref{eq:likelihood})) to
find the best-fitting parameters, except now
$\phi_{\rm th}\left(E_{i}\right)=\phi_{0}+\Delta\phi_{\rm WEP}\left(\Delta\gamma,\;E_{i}\right)+\Delta\phi_{\rm LIV}\left(\eta,\;E_{i}\right)$
and $\sigma_{{\rm tot},i}^{2}=\sigma_{\phi_{{\rm obs},i}}^{2}+\left[\left(\Delta\phi_{\rm WEP}+2\Delta\phi_{\rm LIV}\right)\frac{\sigma_{E_{i}}}{E_{i}}\right]^{2}$.

\begin{figure*}
\vskip-0.2in
\centerline{\includegraphics[angle=0,width=1.0\hsize]{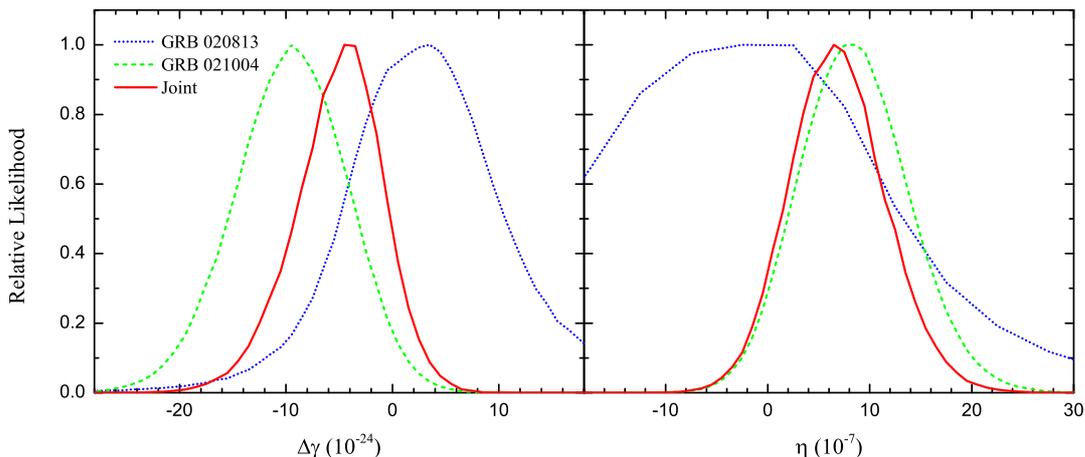}}
\vskip-0.1in
\caption{Individual and joint constraints on both the difference of the $\gamma$ values and the birefringent parameter $\eta$
from the optical polarimetry of GRB 020813 and GRB 021004.}
\label{fig:f5}
\end{figure*}

In Figure~\ref{fig:f5}, we show the marginalized likelihood distributions of $\Delta\gamma$ and $\eta$
derived from the polarimetry of GRB 020813 (dotted curves), GRB 021004 (dashed curves), and their combination
(solid curves), respectively. For the combination of the two bursts, the $3\sigma$ confidence level constraints
on the parameters are $\Delta\gamma=\left(-4.5^{+10.0}_{-16.0}\right)\times10^{-24}$ and $\eta=\left(6.5^{+15.0}_{-14.0}\right)\times10^{-7}$.
With such stringent constraints on both $\Delta\gamma$ and $\eta$, we can conclude that
the birefringence effect arising from violations of both the WEP and Lorentz invariance is insignificant.
These are the first simultaneous verifications of the WEP and Lorentz invariance in the photon sector.

\section{Summary and discussion}
\label{sec:summary}

The WEP states that any two test particles, if emitted from the same astronomical object and traveling through
the same gravitational field, should follow the identical trajectory and undergo the same $\gamma$-dependent
Shapiro delay ($\gamma$ is one of the PPN parameters), regardless of their internal structures (e.g., energies
or polarization states) and compositions. Once the WEP fails, then photons with different circular polarization
states might correspond to different $\gamma$ values, which results in slightly different arrival times for
these two polarization states, leading to birefringence and a frequency-dependent rotation of the polarization
vector of a linear polarized wave. Therefore, linear polarization measurements of astrophysical sources can
provide stringent tests of the WEP through the relative differential variations of the $\gamma$ parameter.
A key challenge in the idea of searching for frequency-dependent linear polarization vector, however, is to
distinguish the rotation angle induced by the WEP violation from any source intrinsic polarization angle
in the emission of photons at different energies. In this work, we simply assume that the intrinsic
polarization angle is an unknown constant, and try to search for the birefringent effect in multiwavelength
polarization measurements of astrophysical sources. By fitting the multiwavelength polarimetric
data of the optical afterglows of GRB 020813 and GRB 021004, we place a statistically robust limit on the WEP
violation at the $3\sigma$ confidence level, i.e., $\Delta\gamma=\left(0.2^{+0.8}_{-1.0}\right)\times10^{-24}$.

As a consequence of LIV, the plane of linear polarization can also generate a frequency-dependent rotation.
Assuming that the rotation angle is mainly caused by LIV, a robust limit on the birefringent parameter $\eta$
quantifying the broken degree of Lorentz invariance can be obtained through the similar fit procedure
in testing the WEP. Using the optical polarimetry of GRB 020813 and GRB 021004, we find that the $3\sigma$
level joint-constraint on the birefringent parameter is $\eta=\left(-0.1^{+1.2}_{-1.7}\right)\times10^{-7}$.

If we consider that the frequency-dependent rotation angle is attributed to violations of both the WEP and Lorentz
invariance, the analysis of the spectropolarimetric data also allows us to simultaneously constrain the difference of
the $\gamma$ values and the birefringent parameter $\eta$. For the combination of GRB 020813 and GRB 021004,
the $3\sigma$ confidence level constraints on both $\Delta\gamma$ and $\eta$ are
$\Delta\gamma=\left(-4.5^{+10.0}_{-16.0}\right)\times10^{-24}$ and $\eta=\left(6.5^{+15.0}_{-14.0}\right)\times10^{-7}$.
While the optical polarimetry of GRBs does not currently have the best sensitivity to WEP tests and LIV
constraints, there is nonetheless merit to the result. First, this is the first time, to our knowledge, that
it has been possible to simultaneously test the WEP and Lorentz invariance through direct fitting of the
multiwavelength polarimetric data of a GRB. Second, thanks to the adoption of multiwavelength polarimetric data set,
our constraints are much more statistically robust than previous results which only with upper limits.
Compared with previous works \cite{2012PhRvL.109x1104T,2013MNRAS.431.3550G,2014MNRAS.444.2776G,2017MNRAS.469L..36Y},
which constrained $\Delta\gamma$ or $\eta$ based on the indirect argument that the relative rotation angle
$|\Delta\phi(E_{2})-\Delta\phi(E_{1})|$ is smaller than $\pi/2$, our present analysis is independent of this argument.
As more and more GRB polarimeters (such as TSUBAME,
COSI, and GRAPE) enter service \cite{2017NewAR..76....1M}, it is reasonable to expect that multiwavelength
polarization observations in the prompt gamma-ray emission of GRBs will be available. Much stronger limits on violations
of both the WEP and Lorentz invariance can be expected as the analysis presented here is applied to larger number of
GRBs with higher energy polarimetry.

It should be noted that the rotation of the linear polarization plane can also be affected by magnetized plasmas
(the so-called Faraday rotation). The rotation angle induced by the Faraday rotation is
$\frac{\Delta\phi_{\rm Far}}{\rm rad}=8.1\times10^{5}\left(\frac{\lambda}{\rm m}\right)^{2}\int_{0}^{L}\left(\frac{B_{\parallel}}{\rm Gs}\right)
\left(\frac{n_{e}}{\rm cm^{-3}}\right)\frac{dL}{\rm pc}$, where $\lambda$ is the wavelength in units of meter,
$B_{\parallel}$ is the magnetic field strength in the intergalactic medium (IGM) parallel to the line-of-sight
(in units of Gauss), $n_{e}$ is the number density of electrons per $\rm cm^{3}$, and $L$ is the distance in units
of pc. Assuming that a cosmic source occurs at $z=2$ (corresponding to a distance of $L\sim10^{10}$ pc),
then for typical IGM with $n_{e}\sim10^{-6}$ ${\rm cm^{-3}}$ and $B_{\parallel}\leq10^{-9}$ Gs,
and at the wavelength of $\lambda\sim10^{-6}$ m where the optical polarimeter operates, we have $\Delta\phi_{\rm Far}\leq10^{-11}$ rad.
It is obvious that the rotation angle $\Delta\phi_{\rm Far}$ at the optical and higher energy band is extremely small,
therefore the Faraday rotation is negligible for the purposes of this work.

\begin{acknowledgements}
This work is partially supported by the National Natural Science Foundation of China
(grant Nos. 11673068, 11725314, and U1831122), the Youth Innovation Promotion
Association (2017366), the Key Research Program of Frontier Sciences (grant Nos. QYZDB-SSW-SYS005
and ZDBS-LY-7014), and the Strategic Priority Research Program ``Multi-waveband gravitational wave universe''
(grant No. XDB23000000) of Chinese Academy of Sciences.
\end{acknowledgements}



\end{document}